\documentclass[pra,twocolumn,a4paper]{revtex4}   	

\usepackage{geometry}                		
\usepackage{graphicx}				
\usepackage{amsmath}
\usepackage{amssymb}
\usepackage{color}
\usepackage{hyperref}

\usepackage{tikz}
\usetikzlibrary{quantikz2}
\usepackage{adjustbox}

\begin{document}
\title{Quantum Normalizing Flows for Anomaly Detection}

\author{Bodo Rosenhahn, Christoph Hirche}
\address{Institute for Information Processing (tnt/L3S), Leibniz Universit\"at Hannover, Germany}


\date{\today}

\begin{abstract}
A Normalizing Flow computes a bijective mapping from an arbitrary distribution to a predefined (e.g. normal) distribution. Such a flow can be used to address different tasks, e.g. 
anomaly detection, once such a mapping has been learned. In this work we introduce Normalizing Flows for Quantum architectures, describe how to model and optimize such a flow and evaluate our method on example datasets. Our proposed models show competitive performance for anomaly detection compared to classical methods, esp. those ones where there are already quantum inspired algorithms available. In the experiments we compare our performace to isolation forests (IF), the local outlier factor (LOF) or single-class SVMs.
\end{abstract}
\maketitle
\section{Introduction}

Anomaly detection is the task to identify data points, entities or events that fall outside a normal range. Thus an anomaly is a data point that deviates from its expectation or the majority of the observations. 
Applications are in the domains of cyber-security \cite{EVANGELOU2020101941}, medicine \cite{FERNANDO202067}, machine vision \cite{BroRud2023}, (financial) fraud detection \cite{Reddy21} or production \cite{RudWan2021a}.
In this work we assume  that only normal data is available 
during training. Such an assumption is valid e.g. in production environments where many positive examples are available and events happen rarely  which lead to faulty examples.
During inference, the model has to differ between normal and anomalous samples. This is also termed
semi-supervised anomaly detection \cite{ruff2020deep}, novelty detection \cite{pmlr-v5-scott09a,Schoelkopf99} or one-class classification \cite{Amer13}.

In this work we will make use of Normalizing Flows \cite{Kobyzev2020} for anomaly detection. A Normalizing Flow (NF) is a transformation of an  arbitrary distribution, e.g. coming from a dataset to a 
 provided probability distribution (e.g., a normal distribution). The deviation from an expected normal distribution can then be used as anomaly score for anomaly detection. A defining property of normalizing flows is the bijectivity, thus a NF can be evaluated as forward and backward path, an aspect which is trivial for quantum gates which can be represented as unitary matrices. Another aspect is that on a quantum computer the output is always a distribution of measurements. This distribution can be directly compared to the target distribution by using a KL-divergence measure for evaluation. In general this step will require sampling.
We would like to raise two aspects why quantum anomaly detection (QAD) can be useful. Firstly,  in combination with quantum machine learning algorithms, QAD can question the quality of the decision, just as a safety net to prevent overconfident or useless decisions \cite{ABDAR2021243}. The second advantage lies in the log2 amount of qubits to represent the data compared to the original representation. E.g. in the experiments, the wine and iris datasets are represented as 12 and 28 dimensional feature vector (details are in the experimental section \ref{SecQFAD}), whereas only 4 and 5 qubits are needed on a quantum device.

We therefore propose to optimize a NF using quantum gates and use the resulting architectures for anomaly detection. In the experiments, we compare the resulting quantum architectures with standard approaches for anomaly detection, e.g. based on isolation forests (IF), local outlier factors (LOF) and one-class support vector machines (SVMs) and show a competitive performance. These methods have been selected since former works 
have already presented  quantum implementations for these variants, or one is in general possible as summarized in section \ref{SecSOAQAD}. 
We additionally demonstrate how to use the Quantum Normalizing Flow as generative model by sampling from the target distribution and evaluating the backward flow. A very recent work in this direction has been presented in \cite{rousselot2024generative}.
For the optimization of the quantum gate order, we rely on quantum architecture search and directly optimize the gate selection and order on a loss function. In our experiments we will use as loss the Kullback-
Leibler (KL) divergence and the cosine dissimilarity.

 Our contributions can be summarized as follows:
\begin{enumerate}
\item We propose Quantum Normalizing Flows to compute a bijective mapping from data samples to a normal distribution
\item Our optimized models are used for anomaly detection and are evaluated and compared to quantum usable reference methods demonstrating a competitive performance. 
\item Our optimized models are used as generative model by sampling from the target distribution and evaluating the backward flow.
\item Our source code for optimization will be made publicly available \footnote{\url{http://www.TBA(after- acceptance)
%tnt.uni-hannover.de/staff/rosenhahn/QNF.zip
}}.
\end{enumerate}

\section{Preliminaries}
In this section we give a brief overview of the quantum framework we use later, a summary on Normalizing Flows, and provide an overview of existing classical and quantum driven anomaly detection frameworks.  Three classical and quantum formalizable methods are later used for a direct comparison with our proposed Quantum-Flow algorithm, namely isolation forests, local outlier factors (LOFs) and single-class SVMs.

\subsection{Quantum Gates and Circuits}
We focus on the setting where our quantum information processing device is comprised of a set of $N$ \emph{logical} qubits, arranged as a quantum register (see, e.g., \cite{10.5555/1206629} for further details). 
Thus we use a Hilbert space of our system $\mathcal{H}\equiv (\mathbb{C}^2)^{\otimes N} \cong \mathbb{C}^{2^N}$ as algebraic embedding. Therefore, e.g., a quantum state vector of a 5-qubit register is a unit vector in $\mathbb{C}^{2^5}=\mathbb{C}^{32}$. We further assume that the system is not subject to decoherence or other external noise. 

Quantum gates are the basic building blocks of quantum circuits, similar to logic gates in digital circuits \cite{Selinger2004a}. According to the axioms of quantum mechanics, quantum logic gates are represented by unitary matrices so that a gate acting on $N$ qubits is represented by a $2^{N}\times 2^{N}$ unitary matrix, a quantum gate sequence comprises of a set of such gates which in return are evaluated as a series of matrix multiplications. A quantum circuit of length $L$ is therefore described by an ordered tuple $(O(1), O(2), \ldots, O(L))$ of quantum gates; the resulting unitary operation $U$ implemented by the circuit is the product 
\begin{equation}
  U = O(L)O(L-1)\cdots O(1).  
\end{equation}  

Standard quantum gates include the Pauli-($X$, $Y$, $Z$) operations, as well as \mbox{Hadamard-}, $\textsc{cnot}$-, $\textsc{swap}$-, phase-shift-, and $\textsc{toffoli}$-gates, all of which are expressible as standardised unitary matrices~\cite{nielsen2010quantum}. The action of a quantum gate is extended to a register of any size exploiting the tensor product operation in the standard way. Even though some gates do not involve additional variables, however, e.g., a phase-shift gate $R_X(\theta)$ applies a complex rotation and involves the rotation angle $\theta$ as free parameter.

\subsection{Normalizing Flows}
 A Normalizing Flow (NF) is a transformation of a provided (simple) probability distribution (e.g., a normal distribution) into an  arbitrary distribution by a sequence of invertible mappings. They have been introduced by Rezende and Mohamed~\cite{nf} as a generative 
model to generate examples from sampling a normal distribution.
\begin{figure}
\centering
  \includegraphics[width=0.5\textwidth]{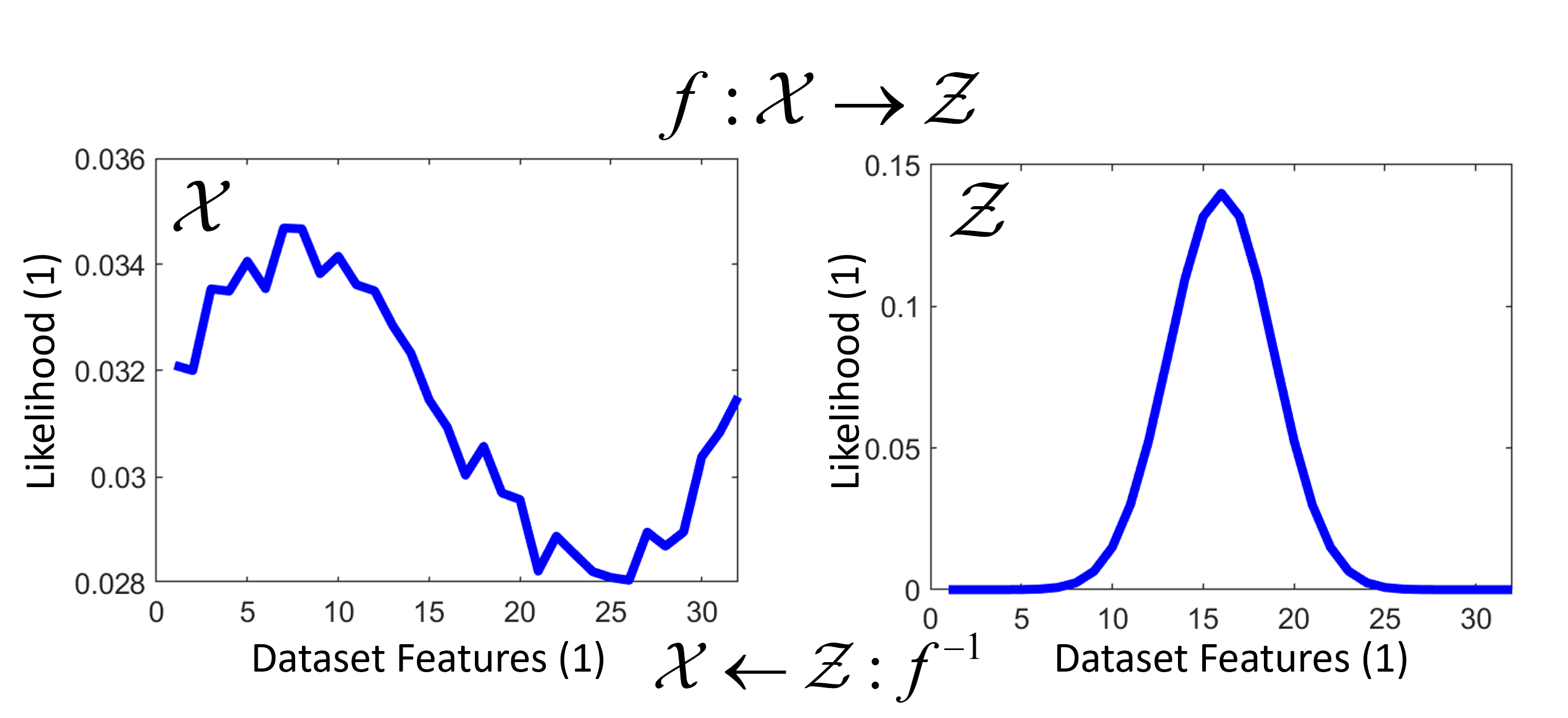}    
\caption{Visualization of a normalizing flow $f$.  It is a bijective transformation $f$ of an  arbitrary distribution (from a given dataset) to e.g. a normal distribution. 
}
\label{fig:NF1}
\end{figure} 
 Compared to other generative models such as variational autoencoders (VAEs) ~\cite{vae} or Generative Adversarial Networks (GANs)~\cite{gan} a NF comes along with the  property that it maps bijectively and is bidirectionally executable. Thus, the input dimension is the same as the output dimension.
Usually, they are optimized via maximum likelihood training and the minimization of a KL-divergence measure. Given two probability distributions $P$ and $Q$, the  Kullback-Leibler-divergence (KL-divergence) or relative entropy is a dissimilarity measure for two distributions,
\begin{eqnarray}
D_{KL}(P\parallel Q)
&=&\sum_{x \in \mathcal{X}}p(x) \log\left( 
\frac{p(x)}{q(x)}\right)
\end{eqnarray}
Since the measure is differentiable, it has been frequently used in the context of neural network models for a variety of downstream tasks such as
image generation, noise modelling,
video generation, audio generation, graph
generation and more \cite{Kobyzev2020}.
Other dissimilarity measures can be based on the cosine-divergence,
optimal transport or the $\chi^2$ measure. Especially 
the cosine-divergence is a useful alternative for quantum computers, as this measure is also directly expressible as quantum gate sequence~\cite{9605277}.
The cosine-dissimilarity of two unit vectors can be expressed as its simple scalar product (denoted as 
$\cdot$),
\begin{eqnarray}
D_{cos}(P,Q)
&=& 1-P \cdot Q
\end{eqnarray}
After the training process, a learned NF can be used in two ways, either as generator or likelihood estimator.
The forward pass $f:\mathcal{X}\rightarrow \mathcal{Z}$ allows for computing the likelihood of observed data points given the target distribution $p_Z$ (e.g. unit Gaussian).
The backward pass $f^{-1}: \mathcal{Z} \rightarrow \mathcal{X}$ allows for generating new samples in the original space $\mathcal{X}$ by sampling from the latent space $\mathcal{Z}$ according to the estimated densities as in~\cite{kingma2018glow, tomINN}.
Figure \ref{fig:NF1} visualizes the change of variables  given the source distribution $p_\mathcal{X}$ and the target distribution $p_\mathcal{Z}$.

Several neural network architectures for such transformations have been proposed in the past~\cite{realnvp, kingma, germain, maf}. Note, that all of these architectures allow for learning highly non-linear mappings, a property which is by definition not possible when solely quantum gates are used (which all comprise linear mappings). We therefore perform later on experiments solely on methods where quantum solutions have been presented. 
Still, since quantum gates are by definition expressable as unitary matrices which are invertible, a quantum gate fulfills the basic properties of an invertible and bijective mapping. Thus, we aim for optimizing a quantum architecture providing the desired Normalizing Flow transformation and use the likelihood estimation from the forward pass to compute an anomaly score for anomaly detection. 

\subsection{Anomaly Detection}
Anomaly detection is the task to identify data points, entities or events that fall outside an expected range. Anomaly detection is applicable in many domains and can be seen as a subarea of unsupervised machine learning. In our setting we focus on a setting where only positive examples are provided which is also termed one-class classification or novelty detection. In the following we will first summarize recent quantum approaches for anomaly detection and will then introduce in more detail three reference methods we will use in our experiments.

\subsubsection{Quantum anomaly detection}
\label{SecSOAQAD}
One of the first approaches to formulate anomaly detection on quantum computers has been proposed in \cite{PhysRevA.97.042315}. 
The authors mainly build up on a kernel PCA and a one-class support vector machine, one of the approaches we will introduce in more detail later. So-called change point detection has been analyzed in \cite{PhysRevLett.119.140506,PhysRevA.98.052305,fanizza2023ultimate}. 
In \cite{PhysRevA.99.052310}, Liang et al. propose anomaly detection using density estimation. Therefore it is assumed that the data follows a specific type of distribution, in its simplest form a Gaussian mixture model.
In \cite{GUO2023129018} the Local Outlier Factor algorithm (LOF algorithm) \cite{LOF2000} has been used and remapped to a quantum formulation. As explained later, the LOF algorithm contains three steps, (a) determine the k-distance neighborhood for each data point $x$, (b) compute the local reachability density of $x$, and (c) calculate the local outlier factor of $x$ to judge whether $x$ is abnormal. 
In \cite{GUO2022127936}
the authors propose an efficient quantum 
anomaly detection algorithm based on density estimation
which is driven from amplitude estimation. They show that their algorithm achieves exponential speed up on the number of training data points $M$ over its classical counterpart. 
Besides fundamental theoretical concepts, many works do not show any experiments on real datasets and are therefore often limited to very simple and artificial examples. Also the generated quantum gate sequences can require a large amount of qubits and they lead to reasonable large code lengths which is suboptimal for real-world scenarios \cite{GUO2023129018, PhysRevA.97.042315}.

In the following we will summarize three classical and well established methods which we will later use for a direct comparison to our proposed Quantum-Flow. For the experiments we used the implementation of these algorithms provided by matlab \cite{MATLAB}. The optimization on the used datasets is very fast and takes less than a second on a standard notebook. 

\subsubsection{Isolation forests}
An isolation forest is an algorithm for anomaly detection which has been initially proposed by Liu et al. \cite{liu2008isolation}. 
It detects anomalies using characteristics of anomalies, i.e. being few and different. The idea behind the isolation forest algorithm is that anomalous data points are easier to separate from the rest of the data. In order to isolate a data point, the algorithm generates partitions on the samples by randomly selecting an attribute and then randomly selecting a split value in a valid parameter range. The recursive partitioning leads to a tree structure and the required number of partitions to isolate a point corresponds to the length of the path in the tree. 
Repeating this strategy leads to an isolation forest and finally, all path lengths in the forest are used to determine an anomaly score.
The isolation forest algorithm computes the anomaly score s(x) of an observation x by normalizing the path length h(x):
\begin{eqnarray}
s(x)&=&2^{-\frac{E[h(x)]}{c(n)}}
\end{eqnarray}
where $E[h(x)]$ is the average path length over all isolation trees in the isolation forest, and $c(n)$ is the average path length of unsuccessful searches in a binary search tree of $n$ observations.

The algorithm has been extended in \cite{Liu2010} and \cite{Talagala2021} to  address clustered and high dimensional data. Another extension is anomaly detection for dynamic data streams using random cut forests, which has been presented in   \cite{pmlr-v48-guha16}. In \cite{Lu2014} quantum decision trees have been proposed, which are the basis for an optional quantum isolation forest. 

\subsubsection{Local outlier factor (LOF)}
The Local Outlier Factor algorithm (LOF algorithm) has been introduced in \cite{LOF2000}. Outlier detection is based on the relative density of a data point with respect to the surrounding neighborhood. It uses the $k$-nearest neighbor and can be summarized as
\begin{eqnarray}
LOF_k(p)&=& \frac{1}{|N_k(p)|}\sum_{o \in N_k(p)} \frac{lrd_k(o)}{lrd_k(p)}
\end{eqnarray}
Here, $lrd$ denotes the local reachability density, $N_k(p)$ represents the $k$-nearest neighbor of an observation $p$.
 The reachability distance of observation $p$ with respect to observation $o$ is defined as
 \begin{eqnarray}
 \tilde{d}_k(p,o)&=&\max(d_k(o),d(p,o))
 \end{eqnarray}
 where
 $d_k(o)$ is the $k$th smallest distance among the distances from the observation $o$ to its neighbors and $d(p,o)$ denotes  the distance between observation $p$ and observation $o$. The local reachability density of observation $p$ is reciprocal to the average reachability distance from observation p to its neighbors.
 \begin{eqnarray}
 lrd_k(p)&=&\frac{1}{\frac{\sum_{o \in N_k(p)}\tilde{d}_k(p,o)}{|N_k(p)|}    }
 \end{eqnarray}
The LOF can be computed on different distance metrics, e.g. an Euclidean, mahalanobis, city block, minkowsky distance or others. In \cite{GUO2023129018} a quantum Local Outlier Factor algorithm (LOF algorithm) has been presented.

\subsubsection{Single Class SVM}

Single class support vector machines (SVMs) for novelty detection have been proposed  in \cite{Schoelkopf99}.  The idea is to estimate a function $f$ which is positive on a simple set $S$ and negative on the complement, thus the probability that a test point drawn from a probability distribution $P$ lies outside of $S$ equals some a priori specified $v$ between $0$ and $1$.
Let $x_i\in \mathbb{R}^N$ denote the training data and
$\Phi$ be a feature map into a dot product space $F$ such that a kernel expression
\begin{eqnarray}
k(x,y) &=& \Phi(x)^T \Phi(y)
\end{eqnarray}
can be used to express a non-linear decision plane in a linear fashion,  which is also known as \textit{kernel trick} \cite{Aizerman67}.
The following formulation optimizes the parameters $w$ and $\rho$ and  returns a function
$f$ that takes the value $+1$ in a \textit{small} region capturing most of the data points, and $-1$
elsewhere.
The objective function can be expressed as quadratic program of the form
\begin{eqnarray}
\textrm{min}_{w\in F, \xi \in \mathbb{R}^t, \rho \in \mathbb{R} }
&& \frac{1}{2} \| w \|^2 + \frac{1}{v^l} \sum_i \xi_i - \rho \\
\textrm{s.t.} &&
(w^T \Phi(x_i)) \geq \rho-\xi_i, \hspace{0.5cm} \xi \geq 0 \nonumber \\
\end{eqnarray}

The nonzero slack variables $\xi_i$ act as penalizer in the objective function. If $w$ and $\rho$ can explain the data, the decision function
\begin{eqnarray}
f(x)&=&\textrm{sgn}((w^T \Phi(x))-\rho) 
\end{eqnarray}
will be positive for most examples $x_i$, while the support vectors in  $\| w \|$ will be minimized, thus the tradeoff between an optimal encapsulation of the training data and accepting outlier in the data is controlled by $v$.
Expanding $f$ by using the dual problem leads to
\begin{eqnarray}
f(x)&=& \textrm{sgn}((w^T \Phi(x))-\rho) \\
&=& \textrm{sgn} \left( \sum_i \alpha_i k (x_i,x)-\rho \right)
\end{eqnarray}
which indicates the support vectors  of the decision boundary with nonzero $\alpha_i$s.
This basic formulation has been frequently used in anomaly detection and established as a well performing algorithm \cite{Amer13,NIPS2007_013a006f}. For more details in linear and quadratic programming we refer to \cite{murty1983,Nguyen2012,Paul09,Bod2023a}.
Quantum SVMs have been matter of research over several years \cite{Heredge2021QuantumSV,9451546,PhysRevLett.113.130503}. 
 An approach for a Quantum one-class SVM has been presented in \cite{PhysRevA.97.042315}.

\section{Method}

In the following we present the proposed method and conducted experiments. The section starts with a brief summary of the used optimization strategy based on quantum architecture search, continues with the evaluation metrics, based on a so-called receiver operating characteristic curve (ROC) and presents the proposed quantum flow anomaly detection framework.
The evaluation is perfomed on two selected datasets, the iris dataset and the wine dataset. Here, we will compare the outcome of our optimized quantum flow with the performance of isolation forests, the LOF and a single class SVM.

\subsection{Optimization}
Optimization is based on quantum architecture search.
 The name is borrowed and adapted from \emph{Neural Architecture Search} (NAS) \cite{Miikkulainen2020,xie2018snas}, which is devoted to the study and hyperparameter tuning of neural networks.
 Many QAS-variants are focussed on discrete optimization and exploit optimization strategies for non-differentiable optimization criteria. In the past,
variants of Gibbs sampling 
\cite{PhysRevResearchLi20}, evolutional approaches  \cite{9870269}, genetic algorithms \cite{RasconiOddi2019,PhysRevLett116.230504} and neural-network based predictors \cite{Zhang2021} have been suggested. A recent survey on QAS can be found in \cite{Zhu23ICACS}.
For this work, we rely on the former work
\cite{RosOsb2023a} proposing Monte Carlo Graph Search. 
The optimized loss function is in our case the Kullback-Leibler divergence, $D_{KL}(P\parallel Q)$ (see Equation (2))  which is the standard loss for many anomaly detection frameworks.  
We will also perform experiments using the cosine-similarity measure. As discussed later, this measure can be evaluated directly from quantum states via a SWAP test. Thus, our approach can be executed on a quantum computer with a sequence comprising of state preparation, the optimized quantum gates sequences and followed by evaluating the quantum cosine dissimilarity as proposed in \cite{9605277}. 
Quantum Architecture search requires at the end (a) a pool of elementary quantum gates $\mathcal{OP}=\{O_1, O_2, \ldots\}$ to sample from, (b) a loss function and (c) some simple hyperparameters, e.g. the maximum length of the quantum gate sequence or a stopping criteria. Then different gate orders are sampled and evaluated. The used quantum architecture search algorithm is based on Monte Carlo graph search  (MCGS) \cite{8632344,Czech_Korus_Kersting_2021}  and measures of importance sampling. MCGS is very efficient, since many operators for quantum computing act locally (e.g. the $H$-Gates, $X$-Gates, $Z$-Gates and more). Therefore many combinations of sampled unitary matrices representing the gate order are algebraically commutative. 
The MCGS-algorithms can be motivated from Monte-Carlo Tree Search (MCTS). It is a heuristic search algorithm for decision processes \cite{8632344}. It makes use of random sampling and balances the exploration-exploitation dilemma in large search spaces. MCTS can be very efficient as it visits more interesting sub trees more often. Thus, it grows asymmetrically and focuses the search time on more relevant parts of the tree. For the MCGS, we
start with the identity operator $ \mathbb{I}$ and build quantum circuits by selecting elementary gates from 
a predefined set $\mathcal{OP}=\{O_1, O_2, \ldots\}$ of elementary quantum gates that we are allowed to apply.
Due to the universality theorem \cite{nielsenQuantumComputationQuantum2000} it is possible to approximate any unitary matrix to arbitrarily good accuracy by using a sufficiently long product of such gates.
The general idea is to grow a graphical model with nodes containing unitary matrices and edges encoding a unitary operator $O_i \in \mathcal{OP}$. The graph is initialized with the identity matrix $ \mathbb{I}$ as root node. An operator $O_i$ is selected and applied to the root node. This yields  a new node by multiplying the selected operator
with the unitary matrix of the parent node (which is the identity matrix in the beginning). If the resulting unitary is already existing as node in the graph, a direct edge from the parent to the already existing node can be added. Else, a new node is generated and connected with the parent node. 
Thus, while growing the graph, the resulting unitary matrices are provided as graph nodes and the underlying quantum code can be computed by finding the shortest path from the root node to the target unitary and by collecting the operators along the edges of the path. 
Figure \ref{fig:QGraphV1} is taken from 
\cite{RosOsb2023a} and shows the general principle.
\begin{figure}
\centering
  \includegraphics[width=0.5\textwidth]{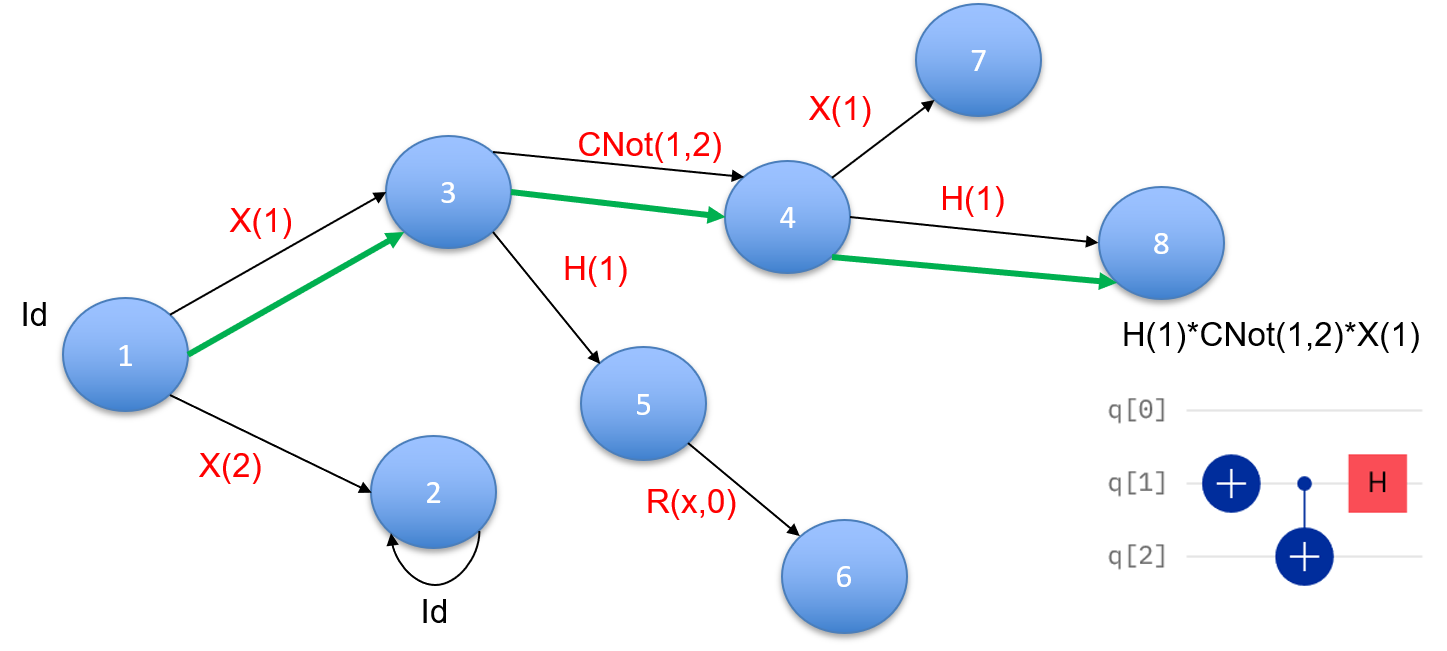}    
\caption{Tiny example graph of quantum circuits. The edges are labelled with elementary gates. The vertices are given by the unitary operator built by taking the product of the gates along the shortest path. (Image taken from \cite{RosOsb2023a})
}
\label{fig:QGraphV1}
\end{figure}
Thus, each node is identified with a possible quantum circuit. Please note, that this graph contains cycles since identical quantum circuits have multiple representations with different gates and gate orders. The remaining challenge is to grow the graph in an efficient manner.
Given a specific task, every node will receive a quality score which is used to compute a probability for its selection and development along this graph area.
 Poisson sampling is then exploited as the underlying sampling process for selecting a vertex to further develop. It is the basic paradigm of
Monte Carlo Search 
\cite{doi:10.1080/01621459.1949.10483310} and adapted Gibbs sampling \cite{george1993variable}  to iteratively grow a graph.
We refer to \cite{RosOsb2023a} for further details.

\begin{figure}
\centering
  \includegraphics[width=0.4\textwidth]{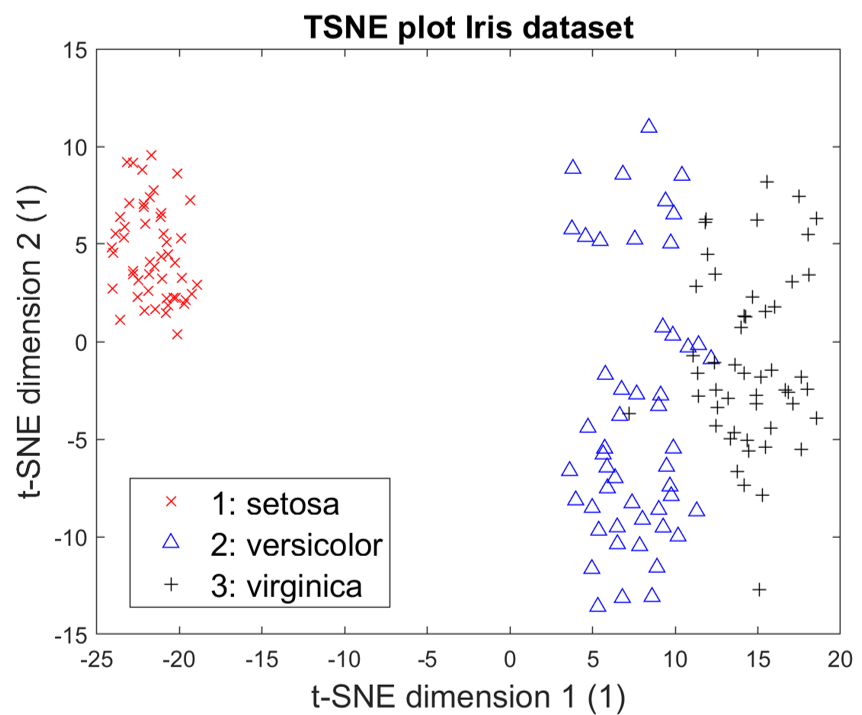}    
   \includegraphics[width=0.4\textwidth]{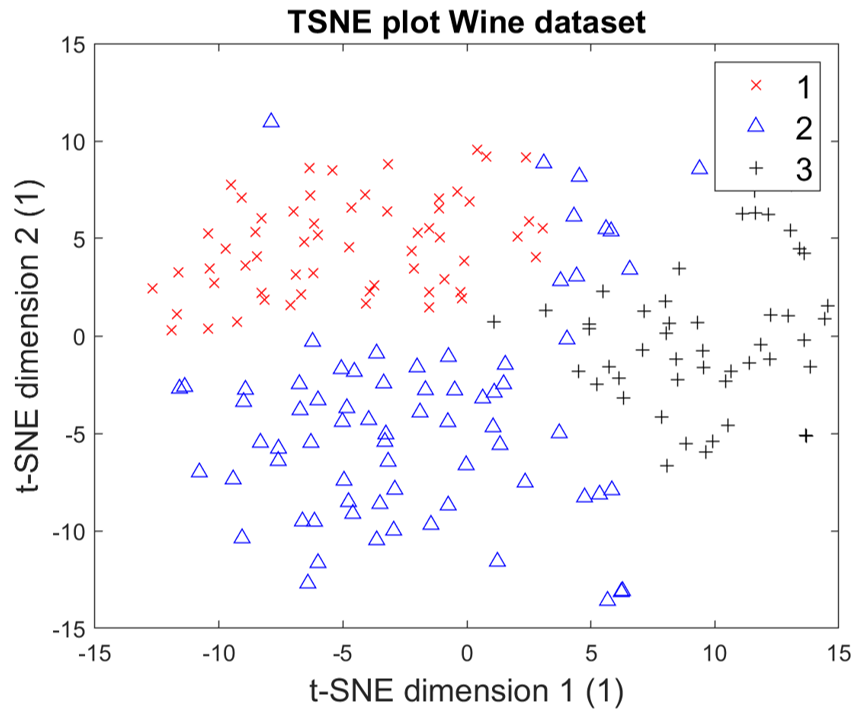} 
\caption{TSNE-plot of the iris (top) and wine (bottom) datasets. The iris dataset has been selected since one class separates very easy from the rest, whereas the remaining classes are more similar and overlapping. The second class of the wine dataset is spreading into the classes one and three.
}
\label{fig:tsnedata}
\end{figure}

\subsection{Evaluation Metrics}
To compare the performance of different algorithms, the area under the ROC-curve (receiver operating characteristic curve) \cite{hanley1982meaning} is a common measure. To summarise, a ROC curve  is a graph showing the performance of a classification model at all classification thresholds. In our case, the classification threshold is the anomaly score of the algorithm and the x-y axes contains the false-positive rate (x-axis) versus the true-positive rate (y-axis) for all possible anomaly thresholds. The area under the generated curve is the AUROC which is 1 in the optimal case when all examples are correctly classified while producing no false positives.
For anomaly detection, the AUROC is the standard measure in the literature for the following reasons, (a) the AUROC is scale-invariant. It measures how well predictions are ranked, rather than their absolute values and (b) the AUROC is classification-threshold-invariant. It measures the quality of the model's predictions irrespective of what classification threshold is chosen.
\begin{figure}[htb]
\centering
  \includegraphics[width=0.45\textwidth]{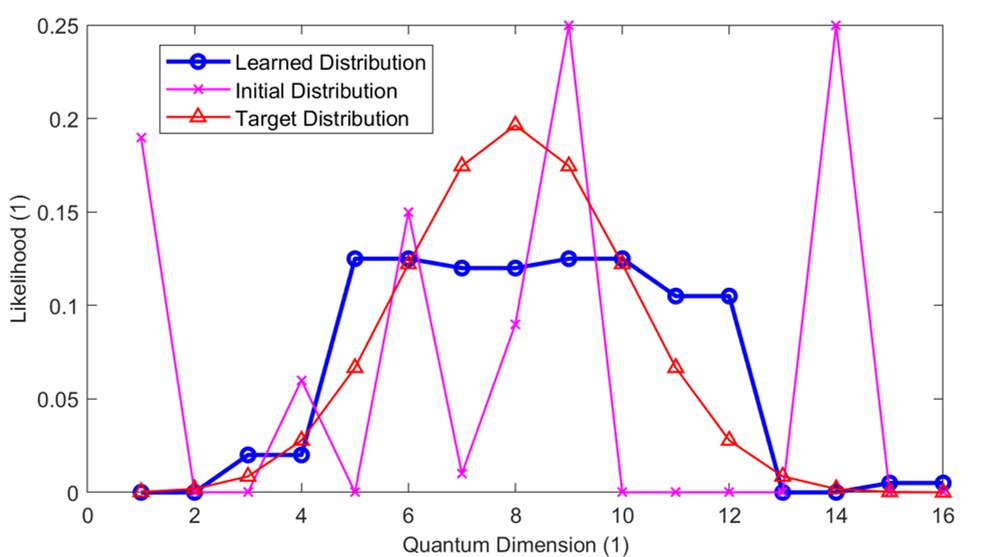} 
\caption{Optimized Normalizing Flow from an input distribution (magenta-cross) to a target distribution (blue-circle) and the comparison to a discrete Gaussian distribution (red-triangle).
}
\label{fig:QDist1}
\end{figure}

\begin{figure}[htb]
\centering
  \includegraphics[width=0.5\textwidth]{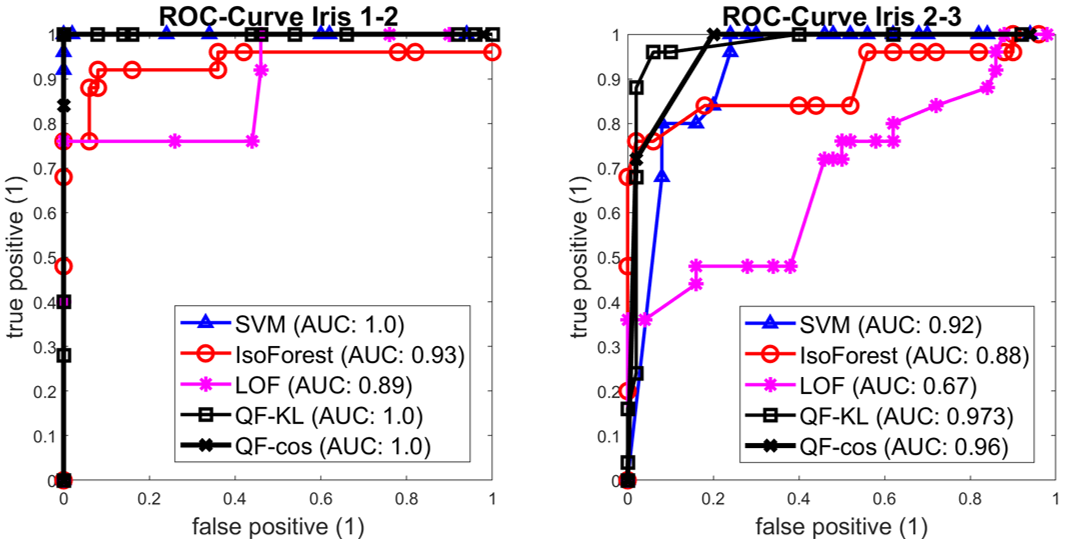}    
   \includegraphics[width=0.5\textwidth]{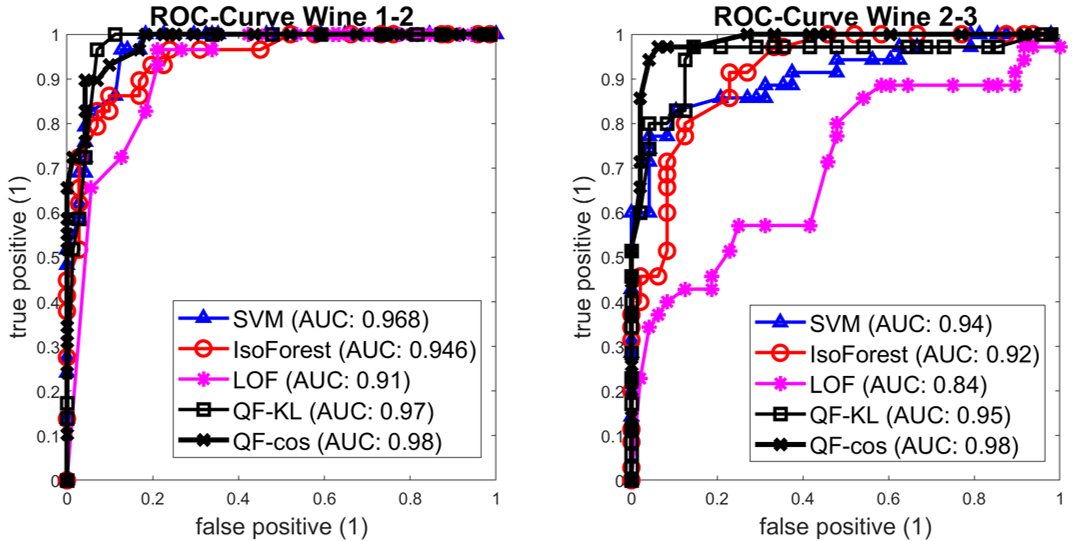}  
\caption{ The resulting ROC-Curves and the aurea-under the ROC-Curve (AUROC) for different settings on the Iris and Wine dataset.  QF-KL shows the performance using the KL-divergence for optimization and QF-cos shows the performance using the cosine dissimilarity score.
}
\label{fig:QGraphV1}
\end{figure}

\subsection{Quantum Flow Anomaly Detection} 
\label{SecQFAD}

\begin{table}
\centerline{
\begin{tabular}{|c|c|c|c|c|c|c|}
\hline
Dataset & Dim &BDim & qubits& \# Train  &\# Test   \\
\hline
Iris 1-2  & 4 & 12 & 4 &25 & 25/50    \\
Iris 2-3 & 4 & 12 & 4 &25 & 25/50    \\
Wine 1-2  & 14 & 28 & 5 &29 & 29/71  \\
Wine 2-3 & 14 & 28 & 5 &35 & 35/48   \\
\hline
 \end{tabular}}
 \caption{Datasets overview, the used normal class, the anomal class and train/test splits.}
 \label{tab:QAD1}
 \end{table} 
 
For the experiments, the classical \textit{wine} and  \textit{iris} datasets were used. The  datasets present multicriterial classification tasks, with three categories for the wine dataset, and three for the iris dataset.  The datasets are all available at the UCI repository \cite{Dua:2019}.
To model an anomaly detection task using a quantum circuit, first the data is encoded as a higher-dimensional binary vector.
Taking the iris dataset as a toy example, it consists of $4$ dimensional data encoding \textit{sepal length}, \textit{sepal width}, \textit{petal length} and \textit{petal width}.
A kMeans clustering on each dimension with $k=3$ is used on the training data. Thus, every datapoint can be encoded in a $4\times 3=12-$dimensional binary vector which contains exactly $4$ non-zero entries. 
The iris dataset contains three categories, samely \textit{setosa, versicolor} and \textit{virginica}.
We select 50\% of data points from one class (e.g. \textit{setosa}) for training and use the remaining datapoints, as well as a second class (e.g. \textit{virginica}) for testing. Thus, the \textit{setosa} test cases should be true positives, whereas the \textit{virginica} should be correctly labeled as anomalies.
Figure \ref{fig:tsnedata} shows a TSNE-plot (t-distributed stochastic neighbor embedding plot) \cite{NIPS2002_6150ccc6} of the iris dataset on the top and a TSNE plot of the wine dataset on the bottom. The iris dataset has been selected since one class separates very easily from the rest, whereas the remaining classes are more similar and overlapping. The second class of the wine dataset is spreading into the classes one and three, here the anomaly detection is also  challenging. These properties will also be reflected in the anomaly scores in the experiments.
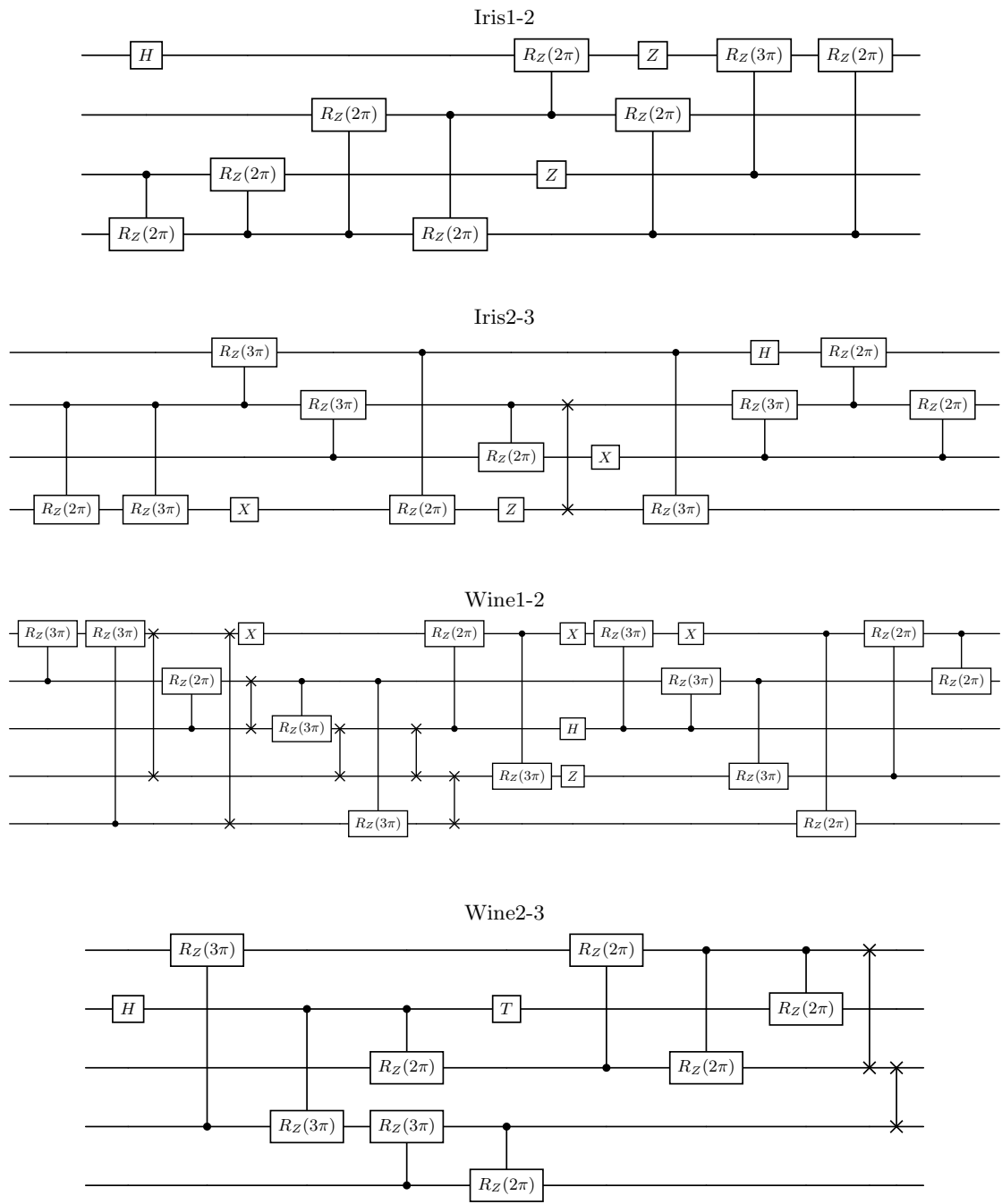
\begin{figure*}
Iris1-2\\[0.1cm]
\begin{adjustbox}{width=0.85\textwidth}
\begin{quantikz}
&\gate{H}&&&& \gate{R_Z(2\pi)} &\gate{Z} &\gate{R_Z(3\pi)}&\gate{R_Z(2\pi)}& \\
&&&\gate{R_Z(2\pi)} & \ctrl{2} &\ctrl{-1} &\gate{R_Z(2\pi)}&&& \\
&\ctrl{1}&\gate{R_Z(2\pi)}&& &\gate{Z}&&\ctrl{-2}&& \\
&\gate{R_Z(2\pi)}&\ctrl{-1}&\ctrl{-2}&\gate{R_Z(2\pi)} &&\ctrl{-2}&&\ctrl{-3}&
\end{quantikz}
\end{adjustbox}
\vspace{0.75cm}\\
Iris2-3\\[0.1cm]
\begin{adjustbox}{width=\textwidth}
\begin{quantikz}
&&&\gate{R_Z(3\pi)}&&\ctrl{3}&&&&\ctrl{3}&\gate{H}&\gate{R_Z(2\pi)}&&  \\
&\ctrl{2}&\ctrl{2}&\ctrl{-1}&\gate{R_Z(3\pi)}&&\ctrl{1}&\swap{2}&&&\gate{R_Z(3\pi)}&\ctrl{-1}&\gate{R_Z(2\pi)}&  \\
&&&&\ctrl{-1}&&\gate{R_Z(2\pi)}&&\gate{X}&&\ctrl{-1}&&\ctrl{-1}&  \\
&\gate{R_Z(2\pi)}&\gate{R_Z(3\pi)}&\gate{X}&&\gate{R_Z(2\pi)}&\gate{Z}&\targX{}&&\gate{R_Z(3\pi)}&&&&
\end{quantikz}
\end{adjustbox}
\vspace{0.75cm}\\
Wine1-2\\[0.1cm]
\begin{adjustbox}{width=\textwidth}
\begin{quantikz}[column sep=0.2cm]
&\gate{R_Z(3\pi)}&\gate{R_Z(3\pi)}&\swap{3}&&\swap{4}&\gate{X}&&&&&\gate{R_Z(2\pi)}&\ctrl{3}&\gate{X}&\gate{R_Z(3\pi)}&\gate{X}&&\ctrl{4}&\gate{R_Z(2\pi)}&\ctrl{1}& \\
&\ctrl{-1}&&&\gate{R_Z(2\pi)}&&\swap{1}&\ctrl{1}&&\ctrl{3}&&&&&&\gate{R_Z(3\pi)}&\ctrl{2}&&&\gate{R_Z(2\pi)}&\\
&&&&\ctrl{-1}&&\targX{}&\gate{R_Z(3\pi)}&\swap{1}&&\swap{1}&\ctrl{-2}&&\gate{H}&\ctrl{-2}&\ctrl{-1}&&&&&\\
&&&\targX{}&&&&&\targX{}&&\targX{}&\swap{1}&\gate{R_Z(3\pi)}&\gate{Z}&&&\gate{R_Z(3\pi)}&&\ctrl{-3}&& \\
&&\ctrl{-4}&&&\targX{}&&&&\gate{R_Z(3\pi)}&&\targX{}&&&&&&\gate{R_Z(2\pi)}&&&
\end{quantikz}
\end{adjustbox}
\vspace{0.75cm}\\
Wine2-3\\[0.1cm]
\begin{adjustbox}{width=0.85\textwidth}
\begin{quantikz}
&&\gate{R_Z(3\pi)}&&&&\gate{R_Z(2\pi)}&\ctrl{2}&\ctrl{1}&\swap{2}&& \\
&\gate{H}&&\ctrl{2}&\ctrl{1}&\gate{T}&&&\gate{R_Z(2\pi)}&&&\\
&&&&\gate{R_Z(2\pi)}&&\ctrl{-2}&\gate{R_Z(2\pi)}&&\targX{}&\swap{1}&\\
&&\ctrl{-3}&\gate{R_Z(3\pi)}&\gate{R_Z(3\pi)}&\ctrl{1}&&&&&\targX{}&  \\
&&&&\ctrl{-1}&\gate{R_Z(2\pi)}&&&&&&
\end{quantikz}
\end{adjustbox}
\caption{Obtained quantum gate sequences after optimization. Despite the competitive performance the resulting codes are reasonable small and efficient.}
\label{fig:QGraphQCodes}
\end{figure*}

For the classically computed quantum architecture search, we compute the discrete distribution $Nhist$ as the normalized histogram of the training dataset as input
 and use a binomial distribution ($p=0.5$) as target.
\begin{eqnarray}
X\sim p(X==k) & \sim & Nhist(k,n) \\
Y\sim  p(Y==k) &\sim &\binom{n}{k}p^k(1-p)^{n-k} \label{Eq:binom}
\end{eqnarray}
The optimization task is to find an ordered set of quantum gates (see Equation (1)) which lead a unitary matrix $U= O(L)$
\mbox{O($L$-1)}$\cdots O(1)$ which minimizes the KL-divergence of the transformed input distribution.
\begin{eqnarray}
  &\underset{U = O(L)O(L-1)\cdots O(1)}{min} \hspace{0.2cm} D_{KL}\left( (|U X|)\parallel Y \right)& \nonumber \\
\end{eqnarray}
Figure~\ref{fig:QDist1} shows an example of how an input distribution is transformed towards a target distribution using an optimized quantum gate sequence $U$. 
The KL-divergence between both distributions is used as anomaly score. Here the ROC-curve evaluates all optional thresholds used for anomaly detection as a scale-invariant measure. This is useful, as it measures how well predictions are ranked, rather than their absolute values.

To make the resulting code more suitable for a quantum implementation, we also optimized the cosine-dissimilarity 
\begin{eqnarray}
  &\underset{U = O(L)O(L-1)\cdots O(1)}{min} \hspace{0.2cm} D_{cos}\left( (|U X|), Y \right)& \nonumber \\
\end{eqnarray}
and evaluated the performance. It turns out, that the obtained results for using a KL divergence and cosine score only deviate up to some noise.

\begin{figure}
\centering
  \includegraphics[width=0.5\textwidth]{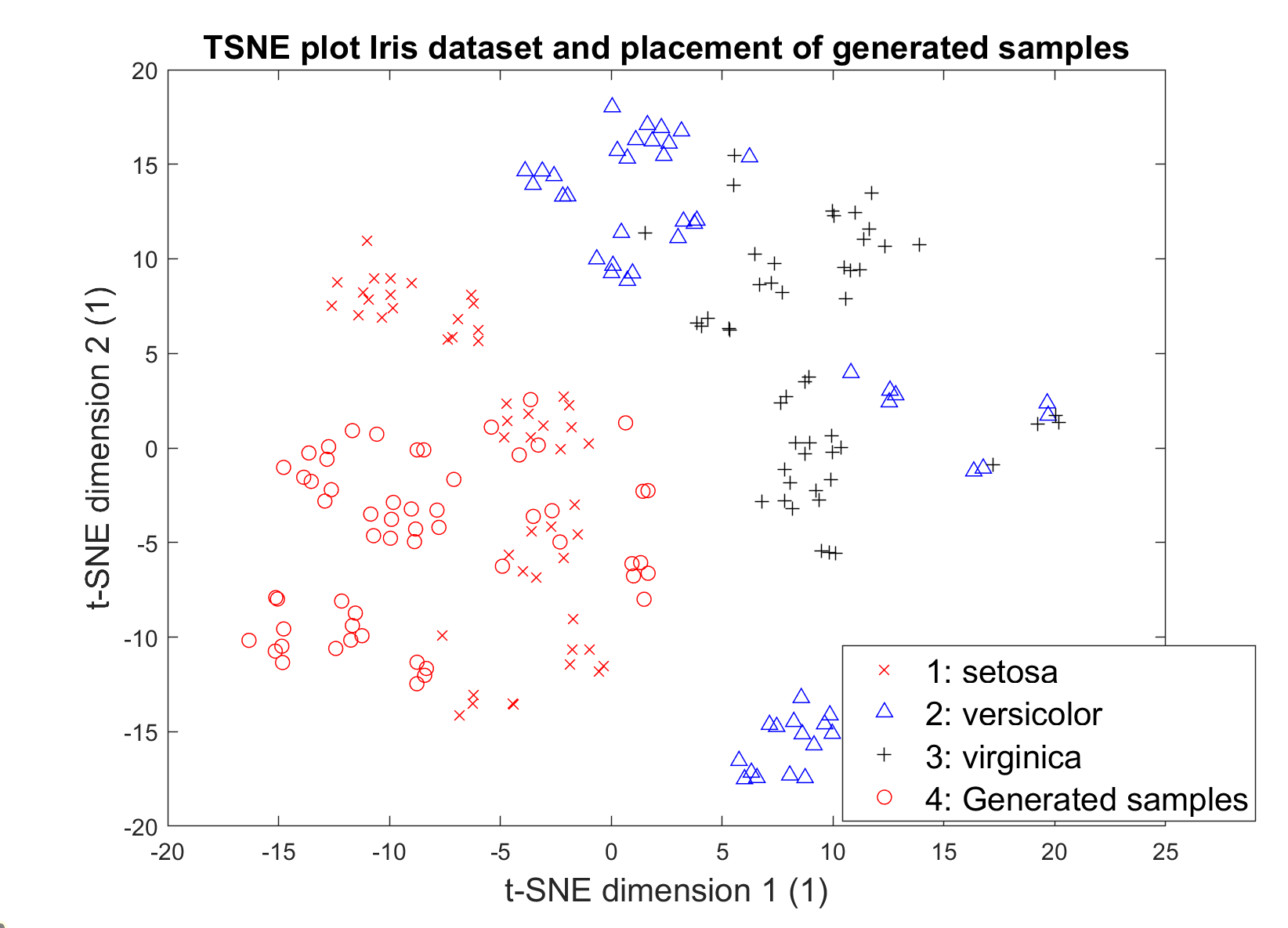}    
   \includegraphics[width=0.45\textwidth]{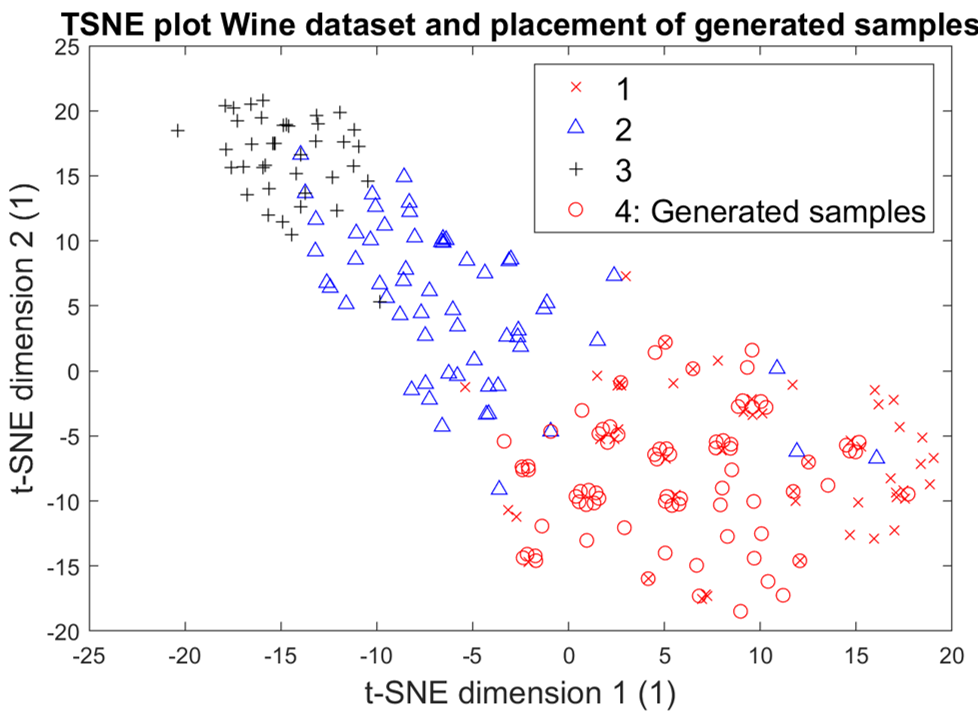}    
\caption{A Normalizing Flow has been trained on the first class for the iris and wine dataset, respectively. The crosses denote the TSNE-plot of the training data. The generated examples  from sampling the normal distribution are shown in circles and consistently fall into the learned domain.
}
\label{fig:QGenExamples}
\end{figure}

 The quantum architecture search algorithm requires as input a pool of optional quantum gates to select from. In our experiments we used the Pauli-($X$, $Y$, $Z$) operations, as well as Hadamard-, $\textsc{cnot}$-, $\textsc{swap}$-, phase-shift-, and $\textsc{toffoli}$-gates. The phase shift gates have a continuous parameter $\theta$ we sampled with 
$\pi, \frac{\pi}{2}, \frac{\pi}{4}$. Thus, only the discrete ordering and selection of quantum gates is optimized by the quantum architecture search.
Table \ref{tab:QAD1} summarizes the used datasets, the settings and the train and test splits. Across all experiments, the optimization of the quantum circuits requires between 10 seconds and one minute. Please note, that this is only the (offline) training stage. Inference on a quantum computer does not require further optimization.
Figure \ref{fig:QGraphV1} summarizes the resulting ROC-Curves and the area-under the ROC-Curve as a final quality measure. Additionally, we provide the ROC-Curves and the obtained results of isolation forests, the local outlier factor (LOF) and the single-class SVM. The final performance is also summarized in table~\ref{tab:QAD2}. It is apparent that our proposed optimized Quantum Flows can achieve competitive performance. Additionally, the code is available as short and highly performant quantum gate sequence. The obtained quantum gate sequences are provided in Figure~\ref{fig:QGraphQCodes}. In the following section we will discuss how the decision algorithm can be implemented on a quantum computer. 

\subsection{Quantum implementation}
The anomaly detection algorithm described in this work essentially consists of two parts. First, we use quantum architecture search to find the unitary that models the normalizing flow. In the second step we evaluate the normalizing flow on new samples to detect a potential anomaly. In thise section we discuss how this second part can also efficiently handled by a quantum computer. 
First, starting from the sample we need to prepare the input to our unitary. A common assumption, see e.g.~\cite[Section III]{PhysRevA.97.042315}, is that we have access to the unitary that prepares the input state $|\Phi\rangle = V |0\rangle$. Note however, that typical classical datasets, such as the examples discussed in this work, often give sparse vectors that can, also without the previous assumption, be encoded efficiently into a pure quantum state~\cite{gleinig2021efficient}. 

Next, we implement the normalizing flow unitary as for the case of our examples given in Figure~\ref{fig:QGraphQCodes}. These consist of elementary gates and the circuit depth is a parameter that can be set in the quantum architecture search. The resulting state is again a pure state, given by $|\psi\rangle = U |\Phi\rangle$. 

The final step is to determine the similarity with the target normal distribution. In general, reading out the exact state of $|\psi\rangle$ can be challenging, e.g. using tomography would require an exponential number of measurements. We hence focus on the cosine-dissimilarity measure. By its definition, it sufficient to compute overlap of $|\psi\rangle$ with the target distribution. To avoid reading out the $|\psi\rangle$, we can instead preprare a state $|\phi\rangle$ according to the normal distribution in Equation~\eqref{Eq:binom} and then apply a simple swap test~\cite{buhrman2001quantum}. 

\begin{figure}
\centering
  \includegraphics[width=0.45\textwidth]{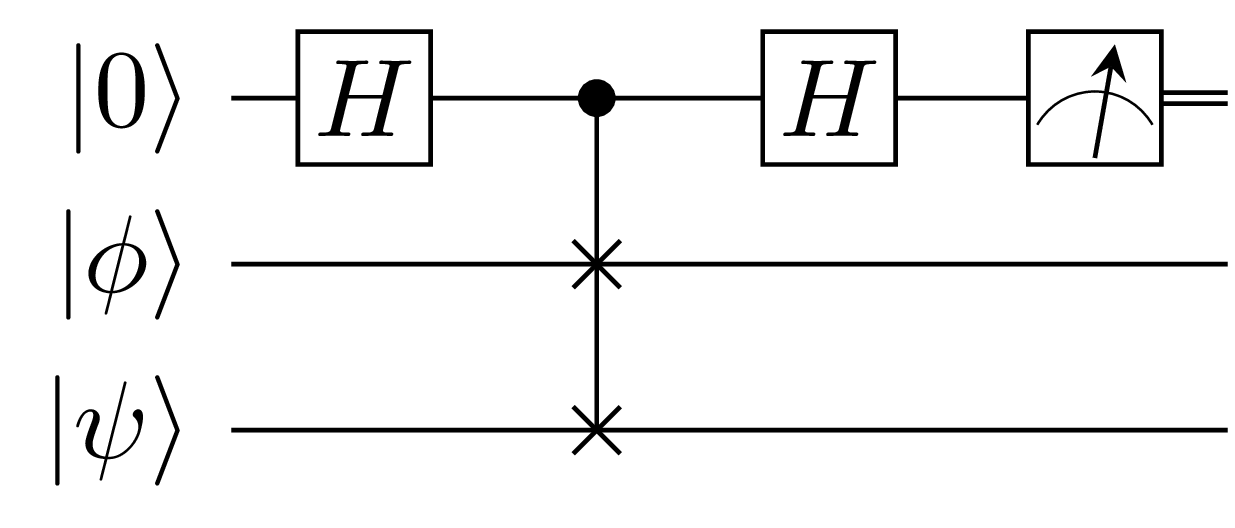}   
\caption{Implementation of the swap test for two quantum states $\psi$ and $\phi$.}
\label{fig:swap}
\end{figure}

This requires one auxilliary qubit, two Hadamard gates, a controlled swap operation and a final quibit measurement. The implementation is shown in Figure~\ref{fig:swap}. The result of the measurement is a binary random variable with probability, 
\begin{align}
    p = \frac12 - \frac12 |\langle \psi|\phi\rangle|^2, 
\end{align}
allowing us to approximate the scalar product $|\langle \psi|\phi\rangle|^2$ up to error $\epsilon$ with $O(\frac1{\epsilon^2})$ samples. 
In summary, we have an algorithm that after initial classical optimization can efficiently run on a quantum computer and reaches higher performance then previous algorithms with known quantum implementations. 

\begin{table}
\centerline{
\begin{tabular}{|c|c|c|c|c|c|}
\hline
Dataset & iso-Forest & LOF& SVM & QF-KL &QF-cos\\
\hline
Iris 1-2  & 0.93 & 0.89& \textbf{1.0} &  \textbf{1.0}   &  \textbf{1.0}\\
Iris 2-3 & 0.88 & 0.67& 0.92 & \textbf{0.97} & 0.96   \\
Wine 1-2  & 0.946 & 0.91& 0.968&  0.97&  \textbf{0.979} \\
Wine 2-3 & 0.92 & 0.84& 0.94 &  0.95&  \textbf{0.98}  \\
\hline
 \end{tabular}}
 \caption{Area under RoC performance on different anomaly detection cases for isolation forests, single-class SVMs and our proposed Quantum Normalizing Flow. QF-KL shows the performance using the KL-divergence for optimization and QF-cos shows the performance using the cosine dissimilarity score.}
 \label{tab:QAD2}
 \end{table}

\subsection{Normalizing Flow as a generator}
As already mentioned in Section 2.2, a Normalizing Flow can be used for several purposes, e.g. anomaly detection as shown in the previous part. Another application is to use the Normalizing Flow as a generator by sampling from the normal distribution and inverting the forward transformation. 
Thus, in the final experiment, a Normalizing Flow has been trained on the first class of both, the iris and wine dataset, respectively. Afterwards, samples from the normal distribution are generated and transformed using the inverse flow which leads to new samples in the original data.
Figure \ref{fig:QGenExamples} visualizes the tsne-plots for these datasets (in crosses) as well as generated examples (in circles) after sampling complex values from the normal distribution and inverting the learned forward mapping $U$. Note, that the backward transformation can lead to non-useful samples since only the distribution on the absolute values is optimized in the forward flow. Thus, sampling complex numbers can lead to unlikely examples.
 We therefore verify the samples by backprojecting them onto the normal distribution again and only select samples which have a small KL divergence. As expected, the generated samples are in the domain of the training data and generalize examples among them.
Figure \ref{fig:QGenExamples} shows the TSNE-data for the three classes and in circles the generated samples which are located around the first class label.

\section{Summary}
In this work, quantum architecture search is used to compute a Normalizing Flow which can be summarized as a bijective mapping from an arbitrary distribution to a normal distribution. The optimization is based on  a Kullback-Leibler-divergence or the cosine dissimilarity. 
Once such a mapping has been optimized, it can be applied to 
anomaly detection by comparing the distribution of quantum measurements to the expected normal distribution. In the experiments we perform comparisons to three standard methods, namely isolation forests, the local outlier factor (LOF) and a single-class SVM. The optimized architectures show competitive performance despite being fully implementable on a quantum computer. Additionally we demonstrate how to use the Normalizing Flow as generator by sampling from a normal distribution and inverting the flow.

\subsection*{Acknowledgments}
This work was supported, in part, by the Quantum Valley Lower Saxony and  under Germany's Excellence Strategy EXC-2122 PhoenixD. We would like to thank Tobias Osborne (Institute for Theoretical Physics at Leibniz University Hannover) for the fruitful discussions and advice on preparing the manuscript.

\bibliographystyle{plain}
\bibliography{refs}

\end{document}